# Aviation sector decarbonization within the hydrogen economy – A UAE case study

Authors: Ghassan Zubi[1], Maximilian Kuhn[1,2], Sofoklis Makridis[3,4], Savio Coutinho[5]

[1]HyStandards GmbH. Email: ghassanzubi@gmail.com.
[2]Hydrogen Europe. Email m.kuhn@hydrogeneurope.eu.
[3]Environmetnal Physics and Hydrogen Technologies Laboratory, Department of Sustainable Agriculture, GR31100, University of Patras, Greece. Email: smakridis@upatras.gr.
[4]Ae4ria Group, Sustainability Unit, Athena Research Centre, Athens, Greece.
[5]Aspire, Advanced Technology Research Council, Abu Dhabi, United Arab Emirates.

**Abbreviations**

AEM: Anion Exchange Membrane
AWE: Alkaline Water Electrolysis
BWB: Blended Wing Body
CCUS: Carbon Capture Utilization and Storage
CSP: Concentrating Solar-thermal Power
EV: Electric Vehicle
GHG: Greenhouse Gas
HFCV: Hydrogen Fuel Cell Vehicle
LCoE: Levelized Cost of Electricity
LGH: Liquid Green Hydrogen
NGCC: Natural Gas Combined Cycle
PEM: Proton Exchange Membrane
PPA: Power Purchase Agreement
R&D: Research & Development
SAF: Sustainable Aviation Fuel
SOE: Solid Oxide Electrolyzer
UAE: United Arab Emirates

**Highlights**

- The development of the UAE power sector is forecasted for the coming decades.
- Large PV farms provide the low-cost electricity needed to produce affordable liquid green hydrogen.
- Liquid green hydrogen in the UAE will reach cost levels competitive with other aviation fuels.
- Hydrogen-powered aviation is undergoing slow development and will become market relevant by 2050.
- Hydrogen as aviation fuel is an opportunity to exploit for the UAE with cost and energy security advantages.

**Abstract**

The UAE aviation sector is vital for its economy and is forecast to grow substantially in the coming decades, increasing thereby fuel consumption. At the same time, the country is committed to cutting greenhouse gas emissions to mitigate climate change. Liquid green hydrogen is expected to emerge as an important aviation fuel in the future. The UAE can use its vast solar energy resources to produce cost-competitive hydrogen at scale, securing so its aviation fuel supply. This development, however, needs several decades to materialize. The PV farms needed to produce the electricity for water electrolysis need yet to be constructed. The infrastructure to produce, liquify, store and transport hydrogen is yet to unfold. Hydrogen-powered aircrafts need to yet evolve from the current small scale demonstration projects to long-haul commercial airplanes. It is realistically by 2050 when hydrogen gains momentum and by 2070 when it becomes the primary aviation fuel. This paper details why and how liquid green hydrogen will find its gap as aviation fuel in the UAE and provides the strategy and policy recommendations to facilitate this development.

1. **Introduction**

The UAE (United Arab Emirates) has a thriving aviation sector which has become a key component of the country's GDP and a major driver of economic growth. For example, Emirates ranked in 2024 among the top 5 airlines in the world in brand value and in number of countries served. Dubai International Airport ranks among the busiest in the world, receiving around 87 million passengers in 2023. The UAE has ambitious projections to further expand its aviation sector. Al Maktoum International Airport, also known as Dubai World Central (DWC), located 30 km southwest of downtown Dubai, is projected to become the world's largest with a capacity for 160 million passengers per year. The airport will be the centerpiece of the massive urban development program Dubai South City and will enhance the connectivity of the UAE as business and trade hub with the rest of the world. At the same time, the UAE has committed itself to reducing GHG (Greenhouse Gas) emissions to net-zero by 2050. The UAE Energy Strategy 2050 provides a framework for environmental obligations while creating a conductive environment for economic growth.

The purpose of this paper is to outline a feasible long-term decarbonization strategy for the UAE aviation sector within the hydrogen economy and provide the policy recommendations for that. The strategy focuses on exploiting technological innovations for the purpose of reducing GHG emissions, while keeping the UAE aviation sector competitive in the international arena and enhancing energy security ahead of a post oil & gas era. The authors recommend a long-term transition to LGH (Liquid Green Hydrogen) as the primary aviation fuel, to be produced in association with PV farms, relying thereby on the vast solar energy resources of the country. Thereby, the UAE will face the challenge of reducing LGH costs to levels competitive with conventional jet fuel. Another major challenge is the development of hydrogen-powered long-haul aircrafts that cover the needs of commercial aviation at scale. This is a global challenge in which the UAE can't set the pace, but still could be a reliable R&D partner, while being among the leading countries in deployment.

Aviation is a challenging sector to decarbonize, as opposed, for instance, to power generation or road transport. Technological transitions in the aviation sector occur generally at slow pace and are subject to strict regulatory procedures. Air travel contributes currently to 2.5% of global GHG emissions, but this share is expected to increase significantly in the coming decades as air traffic intensifies [Klöwer et al., 2018]. The global jet fuel demand in 2023 was 325 million m$^3$, while an average annual growth around 3% can be expected in the coming years. With such trend demand would double by 2050. Hence, finding ways to reduce the carbon footprint of aviation is paramount.

Table 1 categorizes aircrafts by range and share in aviation emissions and indicates the potential alternatives to reduce GHG emissions. Commuter and regional aircrafts have a very small share (4%) in the sectors emissions, two thirds of emissions are attributed to the short and medium haul (1000-3000 km), and 30% to long haul flights. Battery-electric aircrafts have very limited potential to reduce the carbon footprint of the aviation sector. The low gravimetric energy density of batteries imposes serious range limitations, and with that the practical use of such aircrafts, while foreseeable developments are not expected to improve this aspect drastically.

| Table 1: Categorization of aircrafts by range and share in aviation emissions with indication of potential alternatives for GHG reduction | | | | | |
|---|---|---|---|---|---|
| | Commuter | Regional | Short Haul | Medium Haul | Long Haul |
| Seats | 9-50 | 50-100 | 100-150 | 150-250 | >250 |
| Range [km] | <500 | 500-1500 | 1000-2000 | 1500-3000 | 3000-15000 |
| Aviation GHG [%] | <1 | 3 | 24 | 43 | 30 |
| Potential Alternatives | | | | | |
| SAF | Feasible | Feasible | Feasible | Feasible | Feasible |
| Hydrogen | Feasible | Feasible | Foreseeable | Foreseeable | Partly Foreseeable |
| Battery Electric | Feasible | Foreseeable | Not Foreseeable | Not Foreseeable | Not Foreseeable |
| | | | | | |

SAF (Sustainable Aviation Fuel) is emerging as the main short and medium-term solution to address the environmental impact of air travel. SAF is produced from biomass feedstocks such as vegetable oils, tallow, sugar crops, starch crops, agricultural residues, forestry residues, municipal solid waste, etc. Extensive details on this topic are available in the reference [Ali Khan et al., 2023]. Depending on the specific feedstock, deferent processes are used to produce SAF. For instance, sugar and starch crops are processed to ethanol via fermentation, while forestry residues and solid municipal waste are gasified to syngas using the Fischer-Tropsch pathway. A subset of SAF are synthetic aviation fuels, e.g. synthetic kerosene, produced from air-captured carbon and green hydrogen [Colelli al., 2023]. There are several policy frameworks around the global that incentivize SAF, most importantly the Inflation Reduction Act in the USA and the ReFuelEU aviation initiative in the EU. For instance, ReFuelEU prescribes incremental mandates for blending SAF into aviation fuel supplies. For biomass-derived fuels, blends start with a 2% requirement in 2025, increasing to 5% in 2030, and reaching 63% in 2050, while requirements for synthetic kerosene start at 0.7% in 2030 and rise to 28% in 2050.

An advantage of SAF is that it does not require major modifications to existing infrastructure, aircrafts, or engines. This reduces implementation barriers drastically. Accordingly, SAF is expected to have a growing role between today and 2050. Several major airlines have already set SAF targets, most commonly a share of 10% in their fuel consumption in 2030, including American Airlines, Delta Airlines, Lufthansa Group, Ryanair, Cathay Pacific and Qantas Airways, among others.

The perspectives of hydrogen as aviation fuel are promising on the long term, especially for the UAE where green hydrogen can be produced locally from solar power at the needed scale. Furthermore, an aviation sector integrated in the hydrogen economy brings major socio-economic benefits and enhances energy security. Yet, hydrogen-powered aviation is far from mature and is currently limited to demonstrating relatively small aircrafts with modest range. It will take at least until 2040 for hydrogen to have a noticeable role in commercial air transport, while a major impact can be expected by 2050. The path to that point requires major developments in the hydrogen infrastructure, including production, liquefaction, storage and transportation, together with massive upscaling of renewable energy capacities, while key innovations in commercial aircrafts need to materialize, and restructuring and adaptations in airports need to happen.

After this introduction, Section 2 provides an overview on the expected UAE power sector development until 2050 with focus on demand growth, power mix evolution, LCoE (Levelized Cost of Electricity) and carbon footprint. This outline is important to understand the UAE's potential to produce green hydrogen at scale. Section 3 provides an overview on hydrogen production and details the properties of LGH within the context of its potential as aviation fuel and contrasts its foreseeable cost development with that of conventional jet fuel and SAF. Section 4 provides an overview on the innovations and engineering trends in hydrogen-powered aviation and indicates how and when these will eventually translate into commercial aircrafts. Section 5 summarizes the conclusions of this work. Section 6 provides a list of policy recommendations to materialize and accelerate the transition towards hydrogen-powered aviation in the UAE.

2. **Power Sector Decarbonization**

Global warming and climate change are becoming one of the most pressing issues for modern society. The global anthropogenic GHG emissions for 2024 are 59 Gt $CO_2$ eq. [World Data Lab, 2024]. Atmospheric $CO_2$ levels have reached 421 ppm [GML, 2024]. The average earth's surface temperature has already increased by 1.26°C compared to the preindustrial level [ECMWF, 2024]. The Paris Agreement requires the global community to control GHG emissions with the aim to keep global warming at 1.5°-2°C [UNFCCC, 2015]. Global warming has major negative consequences as detailed by the International Panel on Climate Change [IPCC, 2023]. Drastically reducing global GHG emissions requires major changes in the energy sector [Hansen et al., 2019].

A global roadmap towards net-zero emissions by 2050 has been proposed by the International Energy Agency [IEA, 2021]. The strategy calls for massively scaling up PV and wind energy to decarbonize the power sector while intensifying electrification and transitioning to power-to-fuel alternatives. Accordingly, electricity should dominate energy consumption by 2050, with almost 90% of power generation originating from renewables. Internal combustion engine vehicles should be phased out by 2035 in favor of EVs (Electric Vehicles) and HFCVs (Hydrogen Fuel Cell Vehicles). Hydrogen and sustainable fuels should become essential in aviation, shipping, and energy-intensive industries. Furthermore, the growing reliance on low-carbon energy technologies must be intertwined with a dropping energy intensity in GDP, so that the global energy demand by 2050 does not exceed that of today, although serving a much bigger economy. IRENA (International Renewable Energy Agency) has also developed a net-zero roadmap with renewables and energy efficiency being the core elements of decarbonization

[Gielen et al., 2019]. Their cost-benefit analysis shows favorable results for such approach, especially when taking externalities into account. Among others, significant benefits result from avoiding the negative health impact of fossil fuels. These findings are in line with other studies such as [Markandya et al., 2018] and [Vandyck et al., 2018].

Roadmaps for the transition to renewables have multiplied in recent years also through studies of national and regional scenarios, e.g. [Zappa et al., 2019] [Pasaoglu et al., 2018]. For instance, Zubi et al. proposed a power mix for Spain that complements PV and wind with dispatchable power plants, specifically hydropower, CSP (Concentrating Solar Thermal Power) with thermal storage, biogas from waste and energy crops, solid biomass and geothermal energy [Zubi et al., 2009] [Zubi, 2011]. Such a mix can have fossil fuel backup that guarantees firm capacity and that can be used within the existing renewable energy infrastructure, e.g., in hybrid CSP plants, dual fuel gas turbines, solid biomass within co-combustion units, etc.

A global trend in the power sector is a growing reliance on PV and wind power. These impose new challenges as non-dispatchable resources that require forecast. These aspects need to be effectively and efficiently balanced by the power system through flexibilization of dispatchable power plants [EC INEA, 2019] [Gonzalez-Salazar et al., 2018], addition of storage capacities [Zubi et al., 2018] and demand control [Lujano-Rojas et al., 2019] [Battaglia et al., 2017] [Good et al., 2017] [Child et al., 2019]. Significant demand-side management potentials can be unlocked in association with EV charging [Liao et al., 2024] [Kaur & Singh, 2023] [Khamis et al., 2023], power-to-heat applications [Bloess et al., 2018] [Pardo-Garcia et al., 2017], water desalination [Oikonomou & Parvania, 2020] [Al-Nory & El-Beltagy, 2014], etc. The transition to a renewables-based mix comes also with major changes in grid typology and operation [Ismael et al., 2019], and potentially a growing role of distributed generation [Joshi et al., 2021] with the inclusion of new system architectures, including microgrids [Hirsch et al., 2018] [Marzband et al., 2018] and aggregators [Chen & Wu, 2018] [Eid et al., 2016]. Market schemes such as dynamic time-of-use tariffs [Lujano-Rojas et al., 2018] and peer-to-peer trading [Sousa et al., 2019] [Park & Yong, 2017] [Zhang et al., 2018] will gain relevance, while the digitalization of the power system will gradually intensify though advanced metering [Dileep, 2020], internet of thing [Fernández et al., 2023], artificial intelligence [Heymann et al., 2024] [Binyamin et al., 2024] and potentially blockchain [Mengelkamp et al., 2018] [Foti & Vavalis, 2021] [Lazarou & Kotsakis, 2021]. Other foreseeable developments in the power system include refining time and space granularities and modernizing ancillary services to shorten demand-supply distances and facilitate close to real-time trading with the advantage of accurate PV and wind energy forecasting [Najibi et al., 2021] [Parag & Sovacool, 2016].

The here highlighted technological evolution of the power system is fundamental for the UAE as its electricity supply will increasingly rely on solar power in the coming decades. PV will dominate the UAE power mix by the mid of the century as it will have by far the lowest LCoE. The cost advantage of the solar energy path has been emphasized in other studies as well, including [Eveloy & Ahmed, 2022].

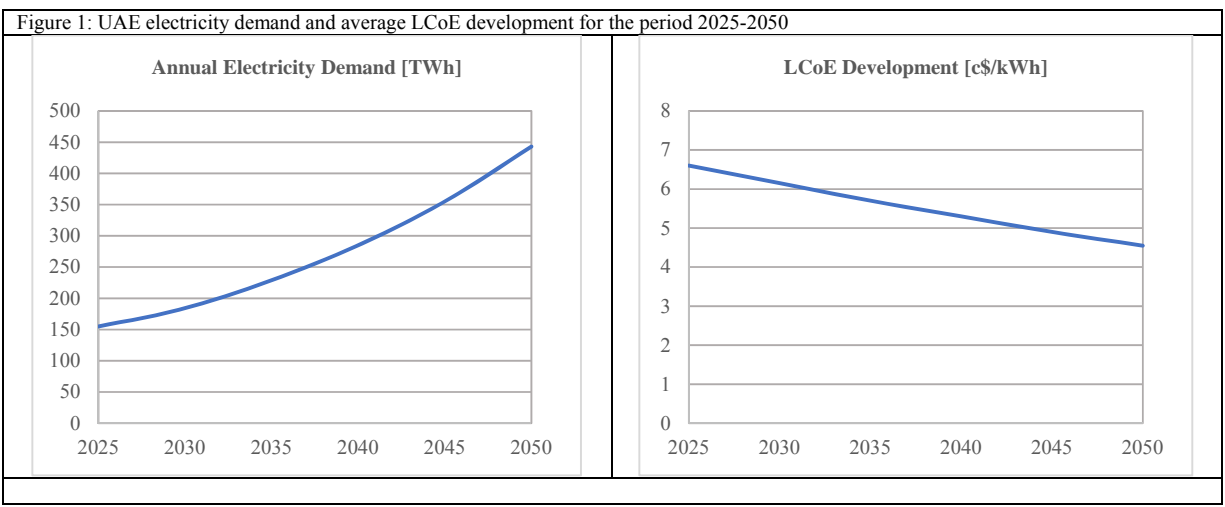

Figure 1: UAE electricity demand and average LCoE development for the period 2025-2050

Figure 1 shows the UAE electricity demand growth for the period 2025 to 2050. On the short term an average annual growth of 3.5% can be anticipated, in line with the development of recent years. For the period 2030-2050, however, a stronger growth of 4.5% on annual average can be expected, among others due to accelerating hydrogen production and growing EV fleet. Hence the annual electricity demand is estimated at 184 TWh in 2030, 285 TWh in 2040 and 443 TWh in 2050. Thereby the average LCoE in power generation is expected to drop from 6.6 c$/kWh in 2025 to 4.5 c$/kWh in 2050 as PV becomes the main electricity source.

The UAE power-mix breakdown for 2030, 2040 and 2050 is illustrated in Figure 2. PV is expected to cover 25% of the annual electricity demand in 2030 and 50% in 2050. Power generation from NGCC (Natural Gas Combined Cycle) power plants is expected to increase in absolute terms but shrink in market share from 43% in 2030 to 31% in 2050. It is assumed that no additional nuclear reactors will be built in the UAE and hence the share of nuclear power in electricity supply will drop from 24% in 2030 to 10% in 2050.

Figure 2: UAE power-mix breakdown for 2030, 2040 and 2050

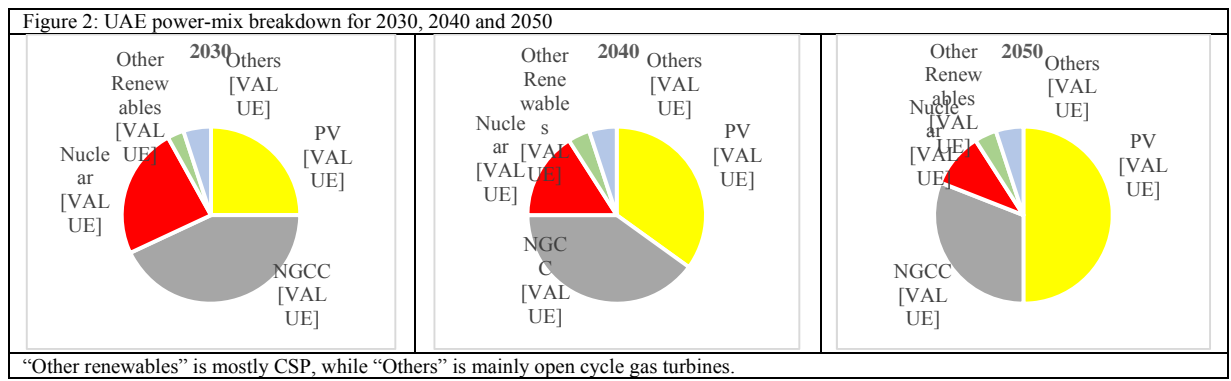

"Other renewables" is mostly CSP, while "Others" is mainly open cycle gas turbines.

As Figure 3 illustrates, the UAE will be able to drastically reduce its specific carbon footprint in the power sector from 280 $gCO_2$/kWh in 2025 to around 100 $gCO_2$/kWh in 2050 with a growing PV role and CCUS (Carbon Capture Utilization and Storage) implementation. This, however, will not result in major changes in the total power sector emissions due to the strong demand growth. The carbon footprint will remain around 45 Mt between today and 2050. Yet, major carbon abatement will result from electrification and power-to-fuel alternatives replacing fossil fuels in transportation and industry.

Figure 3: UAE power sector carbon emissions for the period 2025-2050

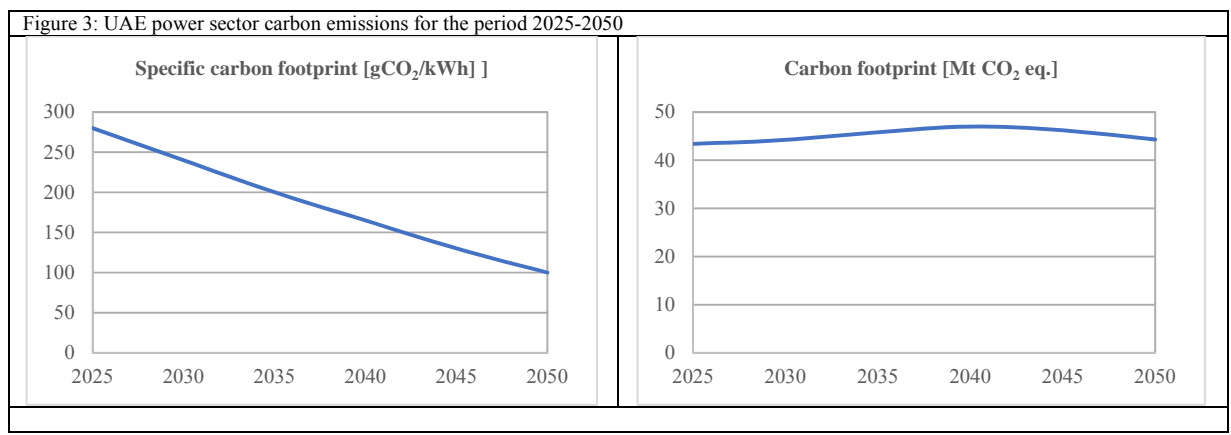

The UAE has already taken and is projecting major steps to decarbonize its power sector. Among others, 4 nuclear reactors have been built at the Barakat plant in Al Dhafra Region, which bring together a nominal power of 5.6 GW, generating around 45 TWh annually. The LCoE of Barakah is 6 c$/kWh, considering thereby 4.5 billion $/GW construction costs, 5% interest rate, 8 years average reactor construction time, 35 years lifetime and 8000 operation hours annually.

Table 2 provides a list of major solar farm projects in the UAE and shows how fast this sector is developing. As clear, there is a strong focus on PV, while CSP plays only a minor role. The UAE has outstanding solar conditions, much available land for large projects [Sgouridis et al., 2016], strong infrastructural and institutional support [Al Naqbi et al., 2019], competitive market and good financing environment [Samour et al., 2022]. While the projects listed in Table 2 bring together an installed PV capacity of 11 GW, there are also 7 GW of PV farms announced by major stakeholders, including TAQA, Masdar, Engie, Helion, and Brooge, to be dedicated to the production of green hydrogen and green ammonia. Rooftop PV systems are also gaining relevance, primarily under the Shams Dubai and the Abu Dhabi Solar Rooftop Program [Alhammami & An, 2021]. The installed PV capacity in the UAE will realistically reach 23 GW in 2030 and contribute with a share of 25% in the annual power supply.

| Table 2: Major solar farms in the UAE | | | | | | |
|---|---|---|---|---|---|---|
| Solar Farm | Type | Location | Owners | Capacity | Complete | Details |
| Mohammed bin Rashid Al Maktoum Solar Park | PV, CSP | Saih Al-Dahal | DEWA | 4660 MW (3960 PV, 700 CSP) | 2030 | Site area: 77 km$^2$<br>Panels: Crystalline Si, CdTe<br>Structure: single axis tracking<br>PPA - Phase 5: 1.7 c$/kWh<br>PPA - Phase 6: 1.63 c$/kWh |
| Al Dhafra Solar Project | PV | Al Dhafra | TAQA<br>Masdar<br>EDF Renew.<br>Jinko Power | 2000 MW | 2023 | Site area: 21 km$^2$<br>Panels: Bifacial crystalline Si<br>Structure: single axis tracking<br>PPA: 1.32 c$/kWh |
| Al Ajban PV Solar Farm | PV | Al Ajban | EWEC<br>EDF Renew.<br>KOWEPO | 1500 MW | 2026 | Panels: Crystalline Si<br>PPA: 1.42 c$/kWh |
| Al Khazna Solar Project | PV | Al Khazna | EWEC<br>EDF Renew.<br>KOWEPO | 1500 MW | 2027 | NA |
| Noor Abu Dhabi Solar Power Plant | PV | Sweihan | TAQA<br>Marubeni<br>Jinko Solar | 1177 MW | 2019 | Site area: 8 km$^2$<br>Panels: Monocrystalline silicon<br>Structure: Fixed<br>PPA: 2.42 c$/kWh |
| Shams Solar Power Station | CSP | Madinat Zayed | Masdar<br>Total Energies<br>Abengoa Solar | 100 MW | 2013 | Site area: 2.5 km$^2$<br>Collectors: Parabolic trough<br>Fossil backup: Natural gas |

PPAs (Power Purchase Agreements) under 2 c$/kWh have become the standard in recent years for large PV farms in the UAE. For instance, Al Dhafra Solar Project has a record PPA at 1.32 c$/kWh. The LCoE of large PV farms will typically remain in the range of 1-2 c$/kWh in the coming decades. This reflects PV system costs around 400 $/kW with PV panel prices in the 100-150 $/kW range and power inverters at around 50 $/kW. Opting for fixed structure or single-axis trackers does not affect the LCoE much. Single-axis trackers are more expensive, but achieve a much better capacity factor, which is close to 25% in the UAE, as opposed to around 19% for fixed struture. The interest rate for solar farm projects in the UAE is realistically in the 4-6% range, and the installation lifetime is 25-35 years.

With the growing share of PV in the power mix, the addition of battery storage will become fundamental. Li-ion batteries of lithium-iron-phosphate chemistry are good candidates for this purpose, while Na-ion batteries are also expected to emerge and take a significant share in this market, potentially with cost advantage and improved circular economy [Metzger et al., 2023] [Soeteman-Hernandez et al., 2023]. The addition of battery-banks allows to make PV farms semi-dispatchable and adapt their output to the daily demand curve, while increasing their role in ancillary services. In line with that, EWEC, for instance, is projecting to add battery storage to the Al Ajban PV Solar Farm. Once the installation of the 1.5 GW PV generator is completed in 2026, EWEC will start adding battery banks to the project, with the first units coming online by 2028.

| Table 3: Energy storage cost for Li-ion battery under different battery cost and cycle life assumptions | | | |
|---|---|---|---|
| | Battery Cost [$/kWh] | | |
| Cycle Life | 200 | 150 | 100 |
| 2000 | 15.8 | 12.1 | 14.2 |
| 3000 | 12.0 | 9.3 | 6.6 |
| 4000 | 10.2 | 7.9 | 5.7 |
| 5000 | 9.1 | 7.1 | 5.1 |
| Assumptions: 90% maximum DoD, 88% roundtrip efficiency (including power converters), 1 cycle/day average usage, 1.5 c$/kWh PV power cost, annual O&M costs at 2% of battery cost and 6% interest rate on investment. | | | |

The cost of power storage depends on several factors, most importantly the battery cost and its cycle life. Table 3 illustrates these costs with assumption that match lithium-iron-phosphate batteries. One can expect large-scale implementation of battery-banks in PV farms to accelerate once storage costs are around 10 c$/kWh, which is likely to be the case by 2030. To match the daily demand curve, a 40% load shift from day to night will be sufficient. Hence, the added cost for making PV semi-dispatchable will be around 4 c$/kWh. On the long run, this cost will drop to around 2 c$/kWh. Hence, the LCoE of semi-dispatchable PV will drop from around 6 c$/kWh in 2030, to around 3.5 c$/kWh in 2050. These values consider carbon penalties of 60 $/t $CO_2$, with specific lifecycle emissions being 70 g $CO_2$/kWh today, dropping to 40 g $CO_2$/kWh in 2050. Hence, PV will become by far the cheapest power source in the UAE.

As illustrated in Figure 2, the UAE will still rely much on natural gas for power generation in the coming decades. The largest power plants in the country are the 8.6 GW Jebel Ali Power and Desalination Plant, the 2.96 GW Fujairah Water and Electricity Generation Complex, the 2.62 GW Al Taweelah, the 2.4 GW Hassyan Power Plant and the 1.67 GW Um Al Nar. The LCoE of NGCC in the UAE is typically around 4.3 c$/kWh today. This considers an initial investment of 600 $/kW, 2 years power plant construction time, 20 years lifetime, 65% capacity factor, 1.7 cS/kWh fuel cost, 25 $/kW O&M costs and 5% interest rate. Should carbon penalties of 60 $/t and specific emissions of 420 g $CO_2$/kWh be considered, then 2.5 c$/kWh additional costs result, bringing the LCoE to around 6.8 c$/kWh. On the long run, NGCC power plants will operate predominantly in association with CCUS, which is competitive against carbon penalties of 60 $/t. Carbon capture and storage costs are in the range of 15-130 $/t at major emissions sites, while afforestation is in the range of 50-250 $/t. On the long run, the production of chemicals and synthetic materials using captured $CO_2$ will create low-cost alternatives to reducing the carbon footprint of power plants [Jiang et al., 2023] [Aresta & Dibenedetto, 2024] [Zhang K et al., 2023]. Taking these factors into account, the average LCoE of NGCC in combination with CCUS will remain around 5.4 c$/kWh in the coming decades. Further information on CCUS can be found in the references [Davoodi et al., 2023] [McLaughlin et al., 2023] [Zhao et al., 2023].

With the average LCoE of the UAE power system declining to 4.5 c$/kWh in 2050 and the specific carbon footprint shrinking to around 100 g$CO_2$/kWh, very favorable conditions are set for hydrogen production at scale. The terms are even better in the case of PV farms dedicated to the production of green hydrogen. Semi-dispatchable PV adapts well for this purpose. This will allow for power supply with a LCoE of 3.5 c$/kWh and specific emissions of 40 g $CO_2$/kWh by 2050 and an annual capacity factor around 65% for the hydrogen production plant.

### 3. Hydrogen as Aviation Fuel

Hydrogen can be obtained from fossil fuels, water or biomass. Accordingly, there are several processes for hydrogen production, including coal gasification [Matamba et al., 2022], steam methane reforming [Ighalo & Amama, 2018], methane pyrolysis [Patlolla et al., 2018], biomass gasification [Rubinsin et al., 2024], biomass pyrolysis [Arregi et al. 2018], biomass fermentation [Akhlaghi & Najafpour-Darzi 2019], water electrolysis [Arsad et al., 2024A], thermochemical water splitting [Ghorbani et al., 202318] and others. Most hydrogen is produced today from natural gas via steam methane reforming without CCUS, what is termed "gray hydrogen" [Rojas et al., 2023]. If this process is complemented with CCUS, then the product is termed "blue hydrogen" [Al Humaidan et al., 2023]. Production via methane pyrolysis gives "turquoise hydrogen" [Diab et al., 2022]. This is a high temperature process in which methane is split with the help of a catalyst (typically metals, metal oxides or carbon) into hydrogen and solid carbon. While hydrogen produced from fossil fuels plays a vital role on the short and medium term, a scalable and sustainable hydrogen economy will eventually rely on water electrolysis as the hydrogen source [Zainal et al., 2024]. This trend can be identified, among others, in the patent landscape of hydrogen production, as recently detailed by Arsad et al. [Arsad et al., 2024B]. Using thereby renewables as power source gives "green hydrogen" [Awad et al., 2024], while using nuclear power gives "pink hydrogen" [Alabbadi et al., 2024]. Further details on the hydrogen "colors" can be found in the reference [Hermesmann & Müller, 2022] [Incer-Valverde et al., 2023A]. An analysis of hydrogen production costs from the different paths is available by [Farhana et al., 2024]. Although green hydrogen has a negligible market share today, it is expected to eventually dominate the hydrogen economy, making it a key component of the net zero pathway [Hassan et al., 2023].

The hydrogen economy is a major component of the global transition to net-zero emissions [Kovač et al. 2021]. Hydrogen has the potential for large-scale implementation across several sectors, including, industry, transportation, and power generation [Sikiru et al., 2024]. Many countries in the Middle East and North Africa have set ambitious strategies and roadmaps for hydrogen production and use [Alsaba et al., 2023] [Razi & Dincer, 2022]. As for today, the hydrogen supply infrastructure is in an initial rollout stage that can serve relatively small capacities [Kim et al., 2023]. These aspects, however, will change in the coming decades with hydrogen production becoming increasingly linked to renewable energy in an accelerating global market [Deloitte, 2023].

Table 4 provides a summary of the key properties of water electroyzers based on the references [Arsad et al., 2023] [Li & Baek, 2021] [Burton et al., 2021] [Motealleh et al., 2021] [Smolinka & Garche, 2021] [El-Emam & Özcan, 2019] [Chi & Yu, 2018] [Acar & Dincer, 2014]. While water electrolysis technologies are continuously under development, two types of electrolyzers dominate currently commercial applications, AWE (Alkaline Water Electrolysis) and PEM (Proton Exchange Membrane), while AEM (Anion Exchange Membrane) and SOE (Solid Oxide Electrolyzer) can be categorized as emerging technologies.

| Table 4: Types of water electrolyzers and their key properties | | | | |
|---|---|---|---|---|
| | PEM | AWE | AEM | SOE |
| Maturity | Commercial | Commercial | Commercial | Early Commercial |
| Charge carrier | $H^+$ | $OH^-$ | $OH^-$ | $O^{2-}$ |
| Electrolyte | Solid polymer | Aqueous solution | Solid polymer | Solid ceramic |
| Working Fluid | Distilled water | Concentrated solution | Distilled water | Steam |
| Anode Material | Pt, Ir, Ru | Ni | Ni alloy | LSMYSZ, $CaTiO_3$ |
| Cathode Material | Pt, Pt-C | Ni alloy | Ni, Ni-Fe, $NiFe_2O_4$ | Nicermets |
| Operation Temperature [C] | 70-90 | 65-100 | 50-70 | 650-1000 |
| Operation pressure [bar] | 15-30 | 2-10 | <35 | <30 |
| Efficiency [%] | 67-84 | 62-82 | ~70 | ~90 |
| Cell voltage [V] | 1.8-2.4 | 1.8-2.4 | ~1.85 | 0.95-1.3 |
| Current density [A/cm$^2$] | 0.6-2 | 0.2-0.4 | 0.1-0.5 | 0.3-1 |

An AWE cell consists of a pair of electrodes separated by a diaphragm, e.g. zircon, and operating in an alkaline solution, commonly potassium hydroxide at 25-30% concentration. Water is split at the cathode in hydrogen and hydroxide anions. The hydrogen is harvested, while the hydroxide anions pass the diaphragm to migrate to the anode and recombine into water and release oxygen. Nickel-based metals are mostly preferred as electrodes due to their affordability and stability. AWE is a mature technology with advantages that include affordability, scalability, durability, and reliance on abundant materials. Disadvantages, on the other hand, are the relatively low current density and low operating pressure, which impose limitations on achieving a compact design. Furthermore, the gas crossover rate is relatively high in such electrolyzers. PEM electrolyzers widely overcome the disadvantages of alkaline electrolyzers. A PEM cell has a solid polymer electrolyte separating the cathode and anode, allowing thereby the flow of hydrogen ions, i.e. protons. These form on the anode, and once passing the membrane, recover their electron on the cathode. The major setback of PEM electrolyzers is the dependance on platinum group metals as catalyst (platinum, iridium, ruthenium, etc.). This results in a cost disadvantage and could cause eventually scalability limitations [Schlichenmaier & Naegler, 2022].

| Table 5: Hydrogen properties | |
|---|---|
| Symbol | H |
| Features | Colorless, odorless, tasteless, flammable gas |
| Atomic number | 1 (1 proton, 1 electron and no neutron) |
| Molar mass [g/mol] | 2.016 |
| Isotopes | Protium ($^1H$), Deuterium ($^2H$), Tritium ($^3H$) |
| Isomers | Ortho-hydrogen, Para-hydrogen |
| Melting temperature [°C] | -259 |
| Boiling temperature at 1 atm [°C] | -253 |
| Critical temperature [°C] | -240 |
| Critical pressure [bar] | 13 |
| Density at 20°C and 1 bar [g/l] | 0.083 |
| Density at 20°C and 350 bar [g/l] | 23.7 |
| Density at 20°C and 700 bar [g/l] | 38.7 |
| Density in liquid phase [g/l] | 70.9 |
| Density of slush hydrogen [g/l] | 86.5 |

| Lower heating value [kWh/kg] | 33.33 |
|---|---|
| Higher heating value [kWh/kg] | 39.39 |
| Latent heat of vaporization [kWh/kg] | 0.128 |
| Ortho- to para-hydrogen conversion heat [kWh/kg] | 0.195 |

A summary of hydrogen properties is provided in Table 5. Hydrogen is the lightest atom with only one proton and one electron. Deuterium (hydrogen with 1 neutron) and tritium (hydrogen with 2 neutrons) are naturally occurring hydrogen isotopes. Hydrogen has very low density at atmospheric pressure and room temperature. Compression to 700 bar allows to reach a density of 38.7 g/l at room temperature, while liquefaction, which requires cryogenic cooling to -253°C, allows to reach a density of 70.9 g/l. The heating value of hydrogen refers to the energy released in its exothermic reaction with oxygen. The higher heating value is 39.39 kWh/kg and results when water is the product, while the lower heating value is 33.33 kWh/kg and results when steam is the product. Hydrogen has two isomers, ortho-hydrogen and para-hydrogen. Ortho-hydrogen is the condition is which both protons of $H_2$ spin in the same direction, while in para-hydrogen they spin in opposite directions. Both have slightly different physical properties. Under atmospheric conditions ortho-hydrogen is in majority with a proportion of 3:1. This does not change significantly when cooling the hydrogen down to -120°C, yet, at lower temperatures the concentration of para-hydrogen increases and reaches 100% at 0°K. Ortho-hydrogen has a higher energy stage and therefore the ortho-para conversion releases heat. Further details on the kinetics of hydrogen liquefaction are available in the reference [Donaubauer et al., 2019].

Figure 4 provides a comparison between hydrogen and conventional fuels in terms of gravimetric and volumetric energy density. The major setback of hydrogen is its low volumetric energy density. This results in major challenges for its use as aviation fuel due to the severe space restrictions in aircrafts.

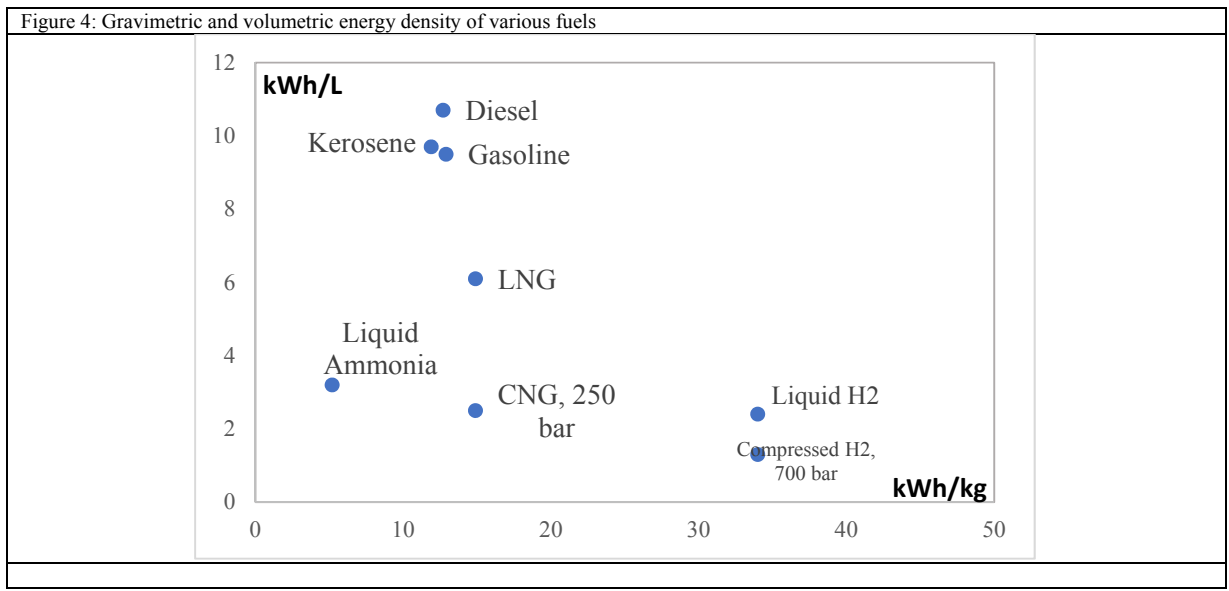

Figure 4: Gravimetric and volumetric energy density of various fuels

Compression is currently the most established hydrogen storage technology with wide implementation in industry and growing use in road transport. Advanced materials are implemented for hydrogen containers, such as carbon fiber-reinforced plastics. A major advantage of compressing hydrogen is that it is energy efficient. This can be done in a near isotherm process in which the gas in chilled and compressed. Chilling allows the process to remain safe as well, especially in the presence of oxygen impurities. The minimum theoretical energy to compress hydrogen isothermally from 20 bar to 700 bar is 1.36 kWh/kg, while the most efficient compressors at industrial scale with hydrogen pre-cooling (e.g. to -40 °C) require around 2 kWh/kg [Tahan, 2022]. Despite its advantages, compressed hydrogen has limited potential as an aviation fuel as its volumetric energy density is about 7 times smaller than that of conventional jet fuel.

Hydrogen liquefaction via cryogenic cooling allows to achieve higher density than its compression. This is an established technology with a variety of processes existing today [Aasadnia & Mehrpooya, 2018], including the Hampson-Linde process [Qin et al., 2023], Collins process [Chen et al., 2022], helium Brayton cycle [Chang et al., 2018] [Nair et al., 2017], magnetic refrigeration [Numazawa et al., 2014], etc. Improving the energy efficiency and reducing system costs are key R&D criteria in this field [Cardella et al., 2023] [Yin & Ju, 2020]. The theoretical minimum energy required for hydrogen liquefaction is 2.7 kWh/kg $H_2$ at a feed pressure of 25 bar. In practice a power consumption of 10 kWh/kg is considered efficient at an industrial scale today, while projections to reduce this value to 6 kWh/kg on the longer run are realistic [Geng & Sun, 2023]. An advantage of liquefaction is that it provides high purity hydrogen, as impurities, most importantly oxygen, condense at higher temperatures. This allows to set and meet hydrogen fuel standards that are fundamental for safety, and optimize operating conditions, especially for degradation-sensitive equipment such as fuel cells [Madhav et al., 2024] [Wang et al., 2014]. Liquid hydrogen can be stored under pressure (cryo-compressed storage) to reduce boil-off losses [Yanxing et al., 2019]. Further cooling of hydrogen to the triple point (-259 °C) leads to slush hydrogen, a mixture of liquid and frozen hydrogen, which has a density of 86.5 g/l. Further information on hydrogen liquefaction, storage and transportation is available in the references [Incer-Valverde et al., 2023B] [Zhang T et al., 2023] [Aziz, 2021] [Andersson & Grönkvist, 2019] [Abdalla et al., 2018].

Figure 5 provides a price contrast between conventional jet fuel, SAF and LGH in the UAE for the period 2030-2070. By 2030, the price of LGH is expected to be around 12 $/kg. This considers power consumption of 52 kWh/kg $H_2$ for water electrolysis and 10 kWh/kg $H_2$ for liquefaction, and power supply from semi-dispatchable PV plants at a cost of 5.5 c$/kWh. The capex and opex are 3.5 $/kg $H_2$ for the water electrolyzers and 4 $/kg $H_2$ for the liquefaction plant. A profit margin of 10% for the LGH producer is considered. By 2070 the price of LGH will drop to around 4 $/kg. This considers power consumption of 47 kWh/kg $H_2$ for water electrolysis and 6 kWh/kg $H_2$ for liquefaction at a power cost of 3.5 c$/kWh, while the capex and opex are 0.8 $/kg $H_2$ for the electrolyzers and 0.9 $/kg $H_2$ for the liquefaction plant. The resulting trend for LGH is equivalent to an average annual cost reduction of 2.8%. On the other hand, the considered jet fuel price is 0.6 $/L in 2030, to add to that carbon penalties of 60 $/t $CO_2$ on the specific emissions of 2.9 kg $CO_2$/L. By 2070, the price of jet fuel is assumed to be 0.8 $/L in a conservative estimate. The resulting contrast between conventional jet fuel and LGH shows cost parity by 2060, while by 2070 LGH will have a 26% cost advantage. On the other hand, SAF and LGH will reach cost parity by 2050, while in 2070 LGH will have a 30% cost advantage.

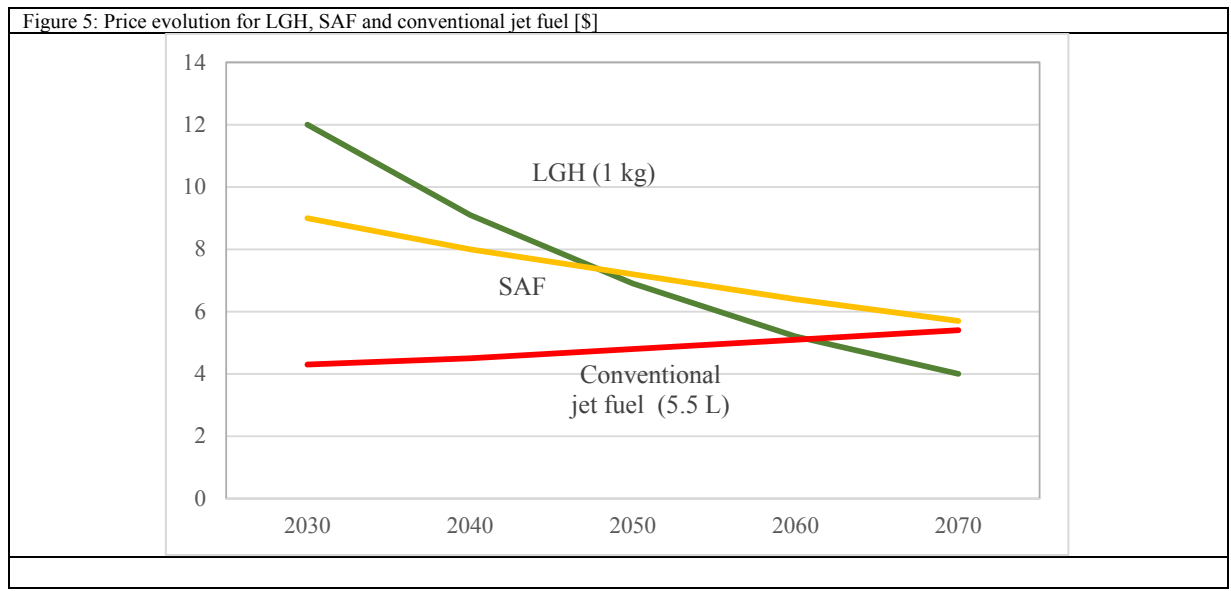

Figure 5: Price evolution for LGH, SAF and conventional jet fuel [$]

Figure 5 assumes hydrogen utilization in fuel cell at 60% efficiency in contrast to jet fuel combustion in gas turbine at 38% efficiency. Hence, 1 kg of hydrogen is equivalent to 5.5 L of jet fuel. Nominal efficiencies around 60% are common in fuel cells used in road transport, while aviation fuel cells lag in this aspect, but are expected to mature to such efficiencies eventually. Figure 5 must also be understood within range restrictions, due to the relatively low specific power of fuel cells. The provided contrast is realistically applicable for regional and short-haul flights. Fuel cells could find implementation in longer range aircrafts, but then rather as part of a hybrid propulsion system together with gas turbines. This reduces the efficiency advantage assumed in Figure 5 and shift the cost parity between LGH and conventional jet fuel towards 2070. In any case, LGH is expected to eventually win the race for the most affordable aviation fuel in the UAE.

The production of hydrogen in association with PV farms at scale is foreseeable in the UAE with several projects announced by major stakeholders, including TAQA, Masdar, Engie, Helion, and Brooge. This trend is positive and fundamental to support the aviation sector decarbonization. LGH will eventually reach a carbon footprint in the range of 2-3 kg $CO_2$ per kg $H_2$ as compared to 16 kg $CO_2$ caused by 5.5 L of conventional jet fuel. This implies around 85% emission reductions by opting for LGH. This said, a major technological transition will be required in the aviation sector to enable the migration from conventional jet fuel to LGH.

## 4. Hydrogen-Powered Aircrafts

The UAE is an intersection of intensive air traffic and has a thriving aviation sector with around 180,000 direct employees. The major airports in the country are listed in Table 6. Dubai International Airport ranks among the busiest in the world, receiving around 87 million passengers in 2023. In terms of fleet size, Emirates has 259 aircrafts, Abu-Dhabi-based Etihad 79, Fly Dubai 79 and the Sharjah-based Air Arabia 58.

Table 6: Major international airports in the UAE

|  | IATA | Location | Airlines | Destinations |
|---|---|---|---|---|
| Dubai International Airport | DXB | Dubai | 84 | 188 |
| Abu Dhabi International Airport | AUH | Abu Dhabi | 58 | 87 |
| Sharjah International Airport | SHJ | Sharjah | 16 | 78 |
| Al Maktoum International Airport | DWC | Jebel Ali | 11 | 40 |

Table 7 summarizes the flight distances from the UAE to the rest of the world. A range of 8000 km reaches a big number of destinations, as it covers the entire Asian, African and European continents. For hydrogen to eventually gain a significant market share, aircrafts that cover this range are required. Flights to North and South America, as well as Australia and New Zealand are within the range of 10,000-15,000 km, yet non-stop flights for such destinations with flight times well over 10 hours are less intensive, i.e. there are less destinations served and less frequent flights.

Table 7: Summary of flight distances from Dubai

| Range | Geographic area | Examples of destinations |
|---|---|---|
| Up to 1000 km | Within Gulf Cooperation Council | Riyadh (867 km), Kuwait City (851 km), Manama (483 km), Muscat (380 km), |
| Up to 8000 km | Asia, Africa and Europe | Tokyo (7926 km), Cape Town (7637 km), Dakar (7610 km), Reykjavik (7361 km), Lisbon (6132 km), Dublin (5925 km) and London (5470 km). |
| Up to 15000 km | North America, South America, Australia and New Zealand | Mexico City (14333 km), Auckland (14203 km) Buenos Aires (13645 km), Los Angeles (13383 km), Sao Paulo (12230 km), Sydney (12050 km), New York (11004 km) and Toronto (11065 km). |
| As for 2024, Emirates flies to 133 destinations of which 50 are in Asia (including 7 Gulf Cooperation Council countries), 38 in Europe, 20 in Africa, 18 in North and South America and 7 in Australia and New Zealand. | | |

There are three major development areas in hydrogen-powered aviation: aircraft design, fuel storage and propulsion system [Tiwari et al., 2024]. Major innovation efforts are required to advance these areas and merge them together. The focus in aircraft design is on the layout optimization to achieve better aerodynamics and fuel economy, while managing the cabin and hydrogen storage space in an efficient distribution that maximizes passenger capacity while allowing to reach a decent flight range. When it comes to fuel storage, compressed hydrogen has limited potential to address the aviation sector needs due to its low volumetric energy density. Liquid hydrogen is notably better in this aspect but requires cryogenic storage. As liquid hydrogen has very different properties than conventional jet fuel, mastering its use in aviation is challenging. Monitoring the temperature, controlling phase change, detecting leakage, ensuring safe release of fuel before emergency landing, etc., pose serious engineering tasks. When it comes to the propulsion system, hydrogen could be used in aircrafts as a combustion fuel in accordingly modified turbines [Zhou et al., 2024], yet it is much more efficient to convert it into electricity in fuel cells and power electric engines.

This said, the current fuel cells on the market have a relatively low specific power, which limits their potential in aviation, especially when it comes to medium and long-haul aircrafts [Fan et al., 2021]. A combination of both, hydrogen gas turbine and hydrogen fuel cell in one aircraft is an option and may become common as the technology matures. More generally, hybrid aircrafts that combine turbines and fuel cells and provide fuel alternatives have gained much attention at R&D level in recent years. Examples of that can be found in the references [Liu et al., 2023] [Seitz et al., 2022] [Seyam et al., 2021] [Collins & McLarty, 2020].

The basic design of commercial aircrafts has not changed much in the past 60 years and remains based on the tube and wing form. This is not due to any inherent advantage but is because aviation prioritizes safety and hence favors well-established layouts while opting rather for gradual improvements. This status, however, is now being challenged by the need for sustainable aviation, where efficiency and alternative fuels become fundamental. Precisely, when it comes to hydrogen, confronting the relatively low volumetric energy density and its impact on the aircraft range are key challenges. This has raised the interest in the BWB (Blended Wing Body) form, where the wings merge with the fuselage of the aircraft, resulting in a streamlined shape with the entire aircraft generating lift while minimizing drag. The improved aerodynamics reduces fuel consumption notably. Furthermore, this layout provides additional and more manageable space for fuel storage. The BWB is a well-established design in military aviation, yet, scaling it up to the size of commercial long-haul aircrafts is challenging in terms of aerodynamic complexity and mechanical load. These factors bring new engineering requirements on the structure and materials. A transition to advanced materials, more specifically light and strong composites such as carbon reinforced plastics, is required to answer these needs. Several important players in the aviation sector are targeting the development of commercial BWB aircrafts, including Airbus, Boing, JetZero and Bombardier, among others. The focus at this stage is on small demonstrators, which could eventually be scaled up in size. Figure 6 provides an impression from exiting development projects.

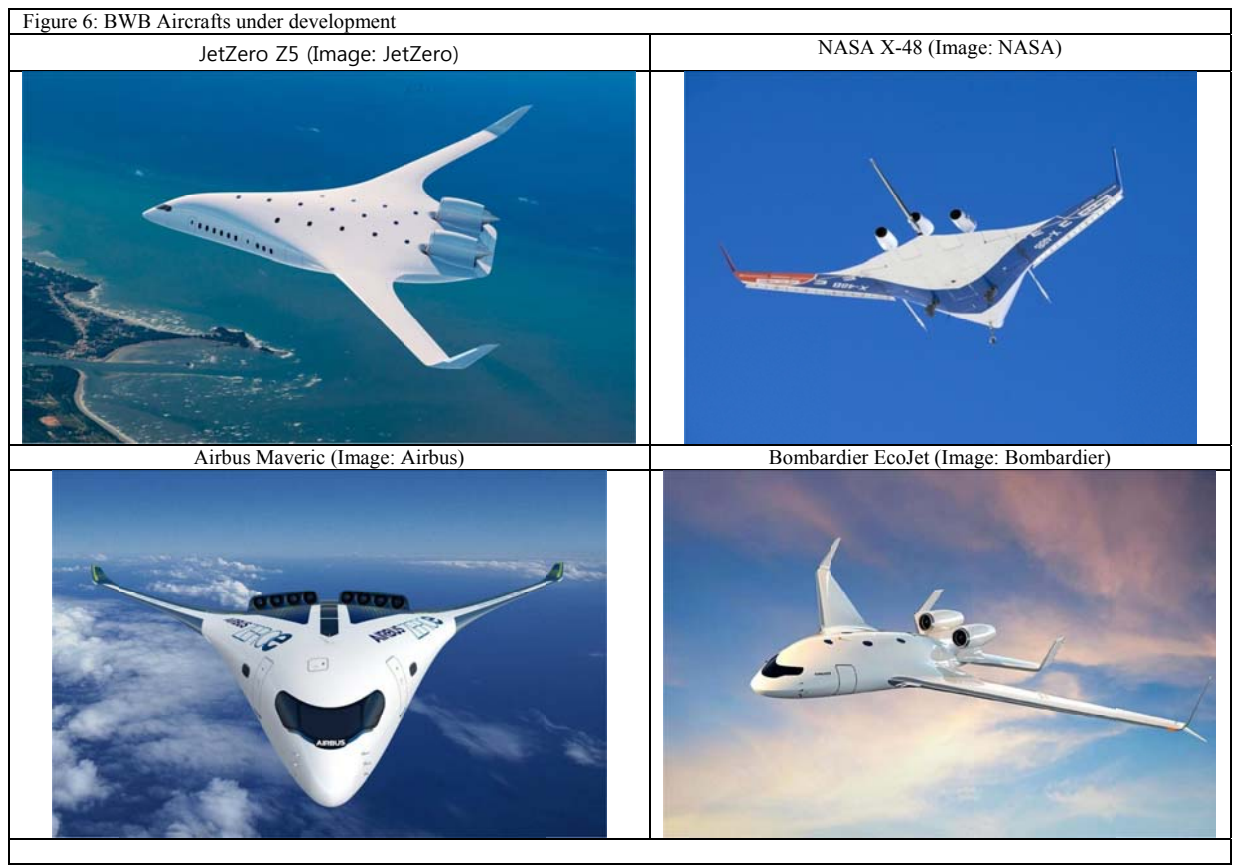

Figure 6: BWB Aircrafts under development

| JetZero Z5 (Image: JetZero) | NASA X-48 (Image: NASA) |
| Airbus Maveric (Image: Airbus) | Bombardier EcoJet (Image: Bombardier) |

As the BWB form is appealing in terms of fuel economy, its development for commercial aviation is relevant independently of its potential for hydrogen. On the other hand, hydrogen-powered aviation is being developed within the context of tube and wing aircrafts as well. A good example of that is Airbus which announced its ZEROe program, with several hydrogen airplane models under development and projected for commercialization by 2035. This includes a tube and wing model with liquid hydrogen storage, combustion in modified gas turbines and turbofan propulsion, with a range of 2000 nautical miles (3704 km) and a capacity for 200 passengers. Furthermore, Airbus has been working on its Maveric aircraft (see Figure 6), a BWB model, initially for fuel economy purposes, but now this model has been added to the ZEROe program. The hydrogen-powered Maveric will eventually compete in the long-haul category.

Boing is interested in the BWB design without committing itself to hydrogen for the moment. Boing has long partnered with NASA in the development of the X-48 (see Figure 6). Initial work on the X-48 started actually by McDonnell Douglas before it was merges with Boing in 1997. By then McDonnell Douglas was losing market share to Airbus and Boing and hence worked on a fuel-efficient aircraft to save its business. Boing continued the work on the technology alongside NASA building thereby subscale models to investigate stability and control. These models have been used to gather data on flight performance to allow the development of new flight control algorithms. The X-48 is projected to evolve into a full-size commercial aircraft with a capacity for 450 passengers.

The California-based company JetZero, on the other hand, is developing its Z5 aircraft (see Figure 6) designed to carry 250 passengers over a range of 5000 nautical miles (9260 km). The Z5 will initially implement existing jet engines and fly on conventional fuel, while transitioning to hydrogen is the long-term vision.

Fuel cell electric propulsion has been successfully demonstrated in small aircrafts, while plans for hybrid systems that integrate hydrogen gas turbines as well are underway. Various fuel cell technologies are being considered for use in aircrafts with the proton exchange membrane and the solid oxide fuel cell being the most promising alternatives [Kazula et al., 2023]. Fuel cell propulsion relies on fully electrical systems, similar to the ones used in battery-electric aircrafts. These systems have major advantages, including zero $NO_x$ emissions, low noise levels and only few moving parts, which can translate into higher reliability and less O&M needs. Fuel cell propulsion has the potential for much better energy efficiency than gas turbines. This aspect, however, is pending of improvements in the coming decades. The efficiency of aviation fuel cells is currently 50% at best, compared to 60% in road transport. Furthermore, a major limitation of current fuel cells is their relatively low specific power, i.e. they are relatively heavy for use in aviation. The current projections indicate that the system-level specific power of fuel cells will reach around 3 kW/kg in 2035, unlocking so the potential for regional and short-haul aircrafts. For medium and long-haul aircrafts, the potential of fuel cells will be initially as part of a hybrid propulsion system.

Several aviation companies are working on fuel cell solutions. The UK-based ZeroAvia has built a 20-seater hydrogen fuel cell-powered aircraft and successfully tested it in 2023. The company is projecting to hit the market within 2025 initially with a regional aircraft of a modest range of 300 nautical miles (556 km) and gradually scale up its technology to eventually commercialize a long-haul BWB aircraft with a capacity for over 200 passengers and a range of 5,000 nautical miles (9260 km) by 2040.

Another important startup is the California-based Universal Hydrogen, which is developing the technology to retrofit aircrafts with fuel cells and electric propulsion. The company has also introduced its LGH refueling strategy with standardized cryogenic tanks that are easy to replace in aircrafts. In 2023 Universal Hydrogen has taken a major step when flying an ATR72 on hydrogen. This is a 60-passenger regional plane with 2 jet engines of 2 MW each. Universal Hydrogen replaced one of these with a fuel cell and electric engine and conducted a test flight, partly with both engines operational and partly cruising exclusively with hydrogen propulsion. The company is aiming to bring its retrofit solution to the market within 2025.

The German company H2FLY, a spinoff from the German Airspace Centre (DLR), is also betting on hydrogen and has built a 4-seater hydrogen fuel cell aircraft, the HY4 (based on the Taurus G4), and conducted its first test flight already in 2016 at Stuttgart Airport. Currently, H2FLY is upgrading its technology from compressed to liquid hydrogen, while developing modular fuel cell systems that can be combined and upscaled to power larger aircrafts. On the short term H2FLY is targeting 20 to 80 seats airplanes for regional air travel. It has recently partnered with Japan Airlines to explore commercial implementation of its technology.

While the developments in aircraft design, hydrogen fuel storage and propulsion system are taking separate paths, these innovations will eventually merge, allowing to overcome the limitations imposed by the low volumetric energy density of hydrogen. Yet, despite the encouraging trend, the evolution of commercial hydrogen-powered aircrafts altogether will be slow and gradual, most importantly in terms of passenger capacity and range, until the long-haul class is properly reached. Realistically, in the coming two decades the focus will be on R&D, regulatory and certification procedures, and initial deployment. Hence, the impact of the technology will be low in terms of market share within this time. Furthermore, considering the lifetime of commercial aircrafts, it will take other 2-3 decades to phase out old-generation aircrafts and modernize fleets. Parallel to that a hydrogen ecosystem that integrates production, liquefaction, transportation, and storage must be built up. Significant adaptations in airports will be required as well. Hence, it is by 2070 that LGH can become the market dominant aviation fuel in the UAE.

5. Conclusions

The UAE aviation sector is poised to grow significantly in the coming decades with ambitious airport development projects, expanding aircraft fleets and intensifying air traffic. This expansion will not focus only on increasing capacity but also on incorporating the latest technologies and sustainable practices. A key issue thereby is the transition to clean aviation fuels. SAF is projected to rapidly gain market share between 2030 and 2050, while LGH is expected to dominate the market on the longer run. A realistic roadmap for SAF and LGH adoption in the UAE aviation sector is suggested in Figure 7. LGH could realistically reach a market share of 50% in 2070, becoming thereby preferential for short and medium-haul flights, while competing with SAF in the long-haul category for flights up to 8,000 km (Asia-Africa-Europe space). To meet this transition, the UAE needs to scale up its LGH production extensively in the coming decades. As detailed in Figure 8, the annual LGH supply to the aviation sector should reach 4.4 Mt by 2070.

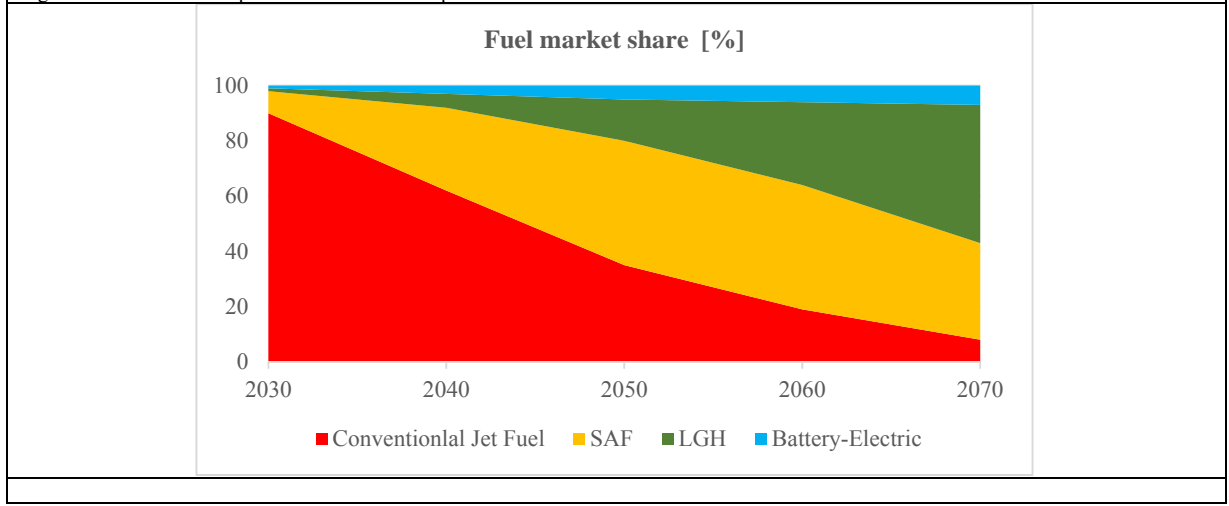

Figure 7: Feasible roadmap for SAF and LGH adoption in the UAE aviation sector

Figure 8: Annual LGH demand by the UAE aviation sector in the coming decades [Mt]

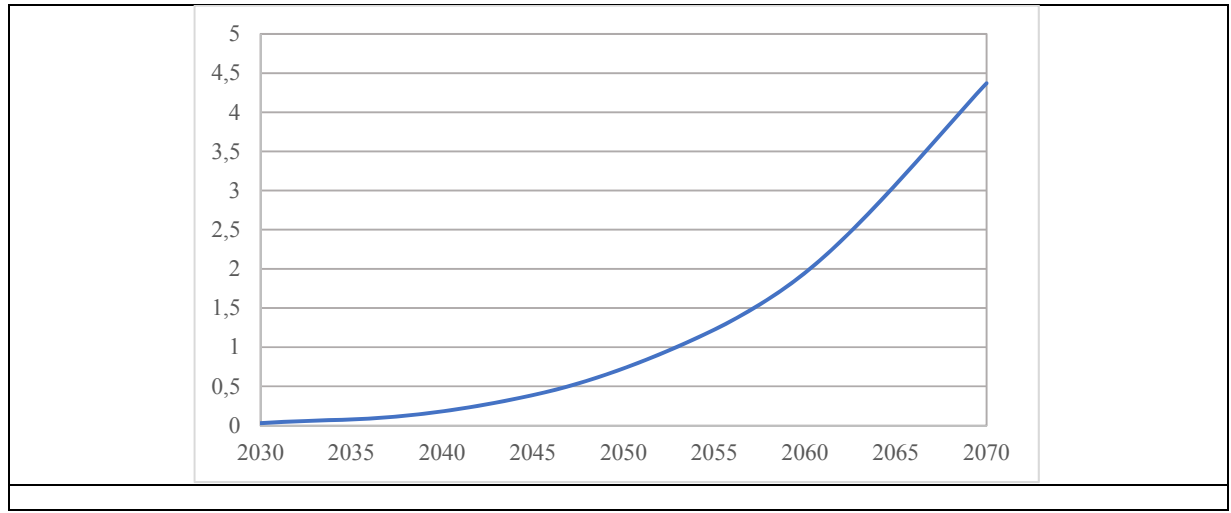

PV will dominate the UAE electricity mix by the mid of the century as it has by far the lowest LCoE while it is by far the largest energy resource. The country's reliance on its vast solar resources not only guarantees a cost-competitive energy supply, but brings huge socio-economic benefits as well, while enhancing energy security and empowering carbon neutrality. The UAE aviation sector can seize this opportunity to secure affordable fuel supply while becoming a global leader in sustainability. LGH can be produced in the UAE in association with semi-dispatchable PV farms. This will allow LGH to reach a carbon footprint comparatively 6 times smaller than conventional jet fuel. On the short term, LGH will be far more expensive than conventional jet fuel and SAF, yet a clear cost advantage over both will eventually crystalize.

The Achilles heel of hydrogen as aviation fuel is its low volumetric energy density. Thereby, there is practically no potential for compressed hydrogen. Liquid hydrogen adapts much better to this role, but even so the volumetric energy density is 4 times less than conventional jet fuel. Hence, hydrogen-powered aircrafts have for a start a clear range disadvantage. Yet, this gap can be reduced through additional fuel-storage space and better fuel economy. Hence, the main development areas to advance hydrogen-powered aviation are the modernization of aircraft design, the integration of cryogenic fuel storage and the reworking of the propulsion system. Furthermore, as LGH has very different properties than conventional jet fuel and SAF, mastering its use in aviation is challenging. Although the cryogenic storage of hydrogen is a well-established technology, its use in aircrafts needs major adaptations to assure efficient use of space, safe handling, fast refueling, reduced boil-off loses, etc.

Major players in the aviation industry are pushing for the development of BWB aircrafts to serve commercial flights. Compared to the tube and wing aircraft, the BWB form creates large internal space while improving fuel economy. This allows for more fuel storage and better specific range and hence provides a counterbalance to the low volumetric energy density of LGH. As for the propulsion system, hydrogen can be combusted in accordingly modified gas turbines. This results in a propulsion system with similar performance to current aircrafts. On the other hand, the use of fuel cells in combination with electric propulsion translates into a significant efficiency advantage, favoring so range and cost. This said, aviation fuel cells must still improve in key aspects before being able to significantly impact air transport. Most importantly a specific power of 3 kW/kg and above and an efficiency around 60% and better should be reached. On the short term, fuel cells will play a modest role in aviation, mostly in regional aircrafts, but as the technology natures to light and efficient systems, it will gradually gain relevance in aircraft engineering, including in hybrid propulsion systems.

Major innovation efforts are required to transform the aircraft design, its propulsion system and fuel storage, and merge these into next generation aircrafts. Developments in these fields are gaining momentum globally with the engagement of big players, including, Airbus, Boing, NASA, Bombardier, JetZero, Universal Hydrogen, ZeroAvia, H2FLY, FlyZero etc. Considering the projections of these major stakeholders it is obvious that the development of commercial hydrogen-powered aircrafts will be slow and gradual, and this will set the pace for the evolution of LGH as aviation fuel. Along this path major developments in the hydrogen infrastructure, including production, liquefaction, storage and transportation will be required.

An important point of discussion is how LGH-powered aviation will eventually affect the cost of air travel. While an accurate quantification is not feasible at this stage, the authors believe that additional costs, if any, would not be prohibitive. Airlines spend roughly 35% of operating revenues on labor costs (pilots, flight crew, office employees, etc.), 30% on fuel supplies, 15% on the purchase and/or lease of aircrafts and their maintenance, 7% on airport fees and 3% are other costs, while the profit margin is typically around 10% of the revenue. Considering the cost breakdown, LGH-powered aircrafts are not expected to alter the total cost of operation dramatically. While in principle the low volumetric energy density of LGH could imply less passenger capacity, in practice there are engineering solutions in the making to marginalize the effects of that, as has been detailed along this work.

The current momentum towards sustainable aviation practices reflects a broader recognition of the urgent need to address climate change. Accordingly, the UAE aviation sector is strengthening its commitment to reducing GHG emissions. This endeavor is evolving into strategies that focus on the adoption of sustainable fuels to improve the environmental performance of the sector. For instance, Emirates has recently established a $200 million Aviation Sustainability Research and Development Fund. As has been detailed in this work, the UAE is well-positioned to embrace LGH as the aviation fuel of the future. Yet, this transition will require a multi-faceted approach, including R&D in hydrogen technologies, demonstration projects, and collaboration with stakeholders to develop the necessary ecosystem for modern aviation.

6. **Policy Recommendations**

LGH-powered aviation in the UAE needs to evolve within an ecosystem that includes the power sector and the hydrogen infrastructure (production, liquefaction, transport, and storage) as well as the aviation fleets and airports. This transformation requires a long-term planning and must engage a big number of stakeholders, including energy companies, airlines, airports, policy makers and regulators, financing institutions, innovation centers, etc. In terms of time frame, the focus between today and 2040 should be primarily on research, development and demonstration of hydrogen-powered aircrafts, while in the 2040s the deployment within the existing aviation ecosystem should be enhanced. After 2050 LGH-powered aviation should be prioritized to cover the growing fuel demand but also to phase out aging fleets. This should allow LGH to eventually dominate the market, while coexisting with other technologies, mostly SAF and to a minor extent regional battery-electric aircrafts. To support the UAE in this transition this work provides the following policy recommendations:

1. **Expanding installed PV capacity:** The UAE should aim to significantly expand its installed PV capacity to around 100 GW by 2050. This will require at least 1000 km$^2$ of land for solar farms, i.e. around 1.2% of the UAE national territory. This ambitious goal underscores the importance of leveraging the UAE's abundant solar energy resources to produce LGH at scale and at a competitive cost to eventually enable a sustainable aviation sector within the hydrogen economy. A resource study to pinpoint the best solar farm locations as well as environmental impact assessments should provide the basis for the land use and licensing. Local and foreign investors should be attracted by simplified regulatory procedures and guaranteed power purchase.
2. **Integrating energy storage solutions into the power system with a circular economy approach:** To achieve a PV share of 50% in the electricity mix by 2050 it is important to upgrade solar farms from non-dispatchable to semi-dispatchable power sources. Most important thereby is to enable a load shift from solar peak hours to the time between sunset and midnight. This requires significant investment in energy storage solutions and should consider not only established technologies such as lead-acid and Li-ion batteries, but also emerging solutions with better environmental performance, security of supply and circular economy perspectives, such as Na-ion batteries. While regulations should facilitate the healthy development of the power storage market, they also should restrict the use of hazardous materials, track the supply chain to discard conflict minerals, and enforce proper collection and disposal schemes, while encouraging reusing and recycling.
3. **Setting credible standards for certifying green hydrogen:** Establishing a robust link between PV farms and hydrogen production facilities is crucial. This requires developing the necessary connecting infrastructure, while facilitating power purchase agreements and ensuring the match between PV power dispatch and electrolyzer operation hours. This link should be established within a generally decarbonizing power sector over time, while phasing out hydrogen production from natural gas. Furthermore, the UAE should strive to set a credible and internationally recognized green hydrogen certification scheme to ensure the global acceptance of its hydrogen as a sustainable fuel.

4. **Addressing the scalability challenges of water electrolysis:** The UAE transition to the green hydrogen economy requires eventually tens of GW of installed electroyzer capacity, in contrast to the few GW of global installed capacity today. It is strategically very important for the UAE to invest in R&D and contribute to the global innovation in this domain to help solve key issues, most importantly securing sustainable material supply and accomplishing circular economy to properly address the scalability challenges. Other key aspects to improve are durability of components and energy efficiency, as well as the need for drastic cost reductions.

5. **Unfolding the liquid hydrogen infrastructure:** Within the coming decades major gradual investments will be required in hydrogen liquefaction plants, cryogenic storage tanks and a transport network. This development will require strong partnerships between the private and the public sectors and between local and foreign investors and must go hand in hand with standards and regulations that guarantee safe production and handling of liquid hydrogen.

6. **Setting the framework for international collaboration in hydrogen Refueling**: Formulating international collaborations and agreements with airlines, airports and countries worldwide is essential for the seamless LGH refueling of UAE airlines in airports abroad and foreign airlines in UAE airports. The UAE should have its voice and role in building up such international and multilateral platform. This can be done within existing associations such as AITA (International Airport Transport Association) and ICAO (International Civil Aviation Organization).

7. **Joining the global efforts in the development of hydrogen-powered aircrafts:** The UAE should thrive to be a key partner in the global development of hydrogen-powered aircrafts and a stakeholder at IP level. Hence, substantial investments in R&D in this domain are critical. Applied research can be encouraged by funding technology incubators, supporting startups, facilitating innovation grants etc. Key areas thereby include novel aircraft designs and composite materials, onboard cryogenic storage, aviation fuel cells and advanced electric propulsion, among others.

8. **Setting standards for hydrogen fuel**: Establishing comprehensive international standards for liquid hydrogen as aviation fuel, including storage and transportation norms, aircraft refueling practices, safety measures, etc., is vital for its widespread adoption. The UAE should support the development of such standards in a multilateral approach and make them part of its national regulations.

9. **Implementing a carbon penalty scheme**: Introducing and enforcing an ambitious carbon penalty scheme favors green hydrogen and encourages stakeholders in the aviation sector to plan ahead and take decisions and measures in favor of this transition. Carbon penalties should be tailored to eventually impact market shares in favor of green technologies and should be based on a comparative techno-economic assessment of the competing alternatives, i.e. conventional jet fuel versus SAF versus LGH.

10. **Addressing critical materials**: Identifying the critical materials required for the transition to LGH-powered aviation is necessary. A dedicated study to quantify material bottlenecks and supply risks and determine their cost impact should be performed and updated every few years as the market grows and technologies advance. Based on that, the UAE should explore alternative materials and develop mitigation strategies to ensure the sustainability and resilience of its aviation fuel supply chain.

By implementing these policy recommendations, the UAE can position itself as a leader in the global transition towards sustainable aviation, leveraging thereby its unique resources and strategic position, while fostering innovation and collaboration in the pursuit of reducing GHG emissions. This ensures the healthy growth of the UAE aviation sector with all its benefits to the national economy, including revenue, direct and indirect jobs, tourism as well as the growing role of the UAE as an international business hub. With this the UAE would also be making a major step towards long-term energy security with national dependance on oil & gas declining and reliance on solar power becoming the backbone of energy supply.